\documentclass[english]{article}
\usepackage[T1]{fontenc}
\usepackage[latin1]{inputenc}
\usepackage{graphicx}
\usepackage{amsmath}
\usepackage{amssymb}
\usepackage[colorlinks]{hyperref}

\begin{document}

\begin{center}
{\bf Anomalous single production of the fourth generation charged leptons at future ep colliders} \\
\end{center}

\vspace{.04cm}

\begin{center}
\emph{A. K. ~\c{C}ift\c{c}i$^{1,*}$, R. ~\c{C}ift\c{c}i$^{2}$, H.~Duran ~Y{\i}ld{\i}z$^{3,\dagger}$, S. ~Sultansoy$^{4,5}$}
\end{center}

\vspace{.02cm}
\begin{center}
\emph{$^{1)}$} Physics Department, Faculty of Sciences, Ankara University, 06100 Tandogan, Ankara, Turkey,

\emph{$^{2)}$} A. Y\"{u}ksel Cad. 7/15, 06010 Etlik, Ankara, Turkey,

\emph{$^{3)}$} Physics Department, Faculty of Sciences and Arts, Dumlup{\i}nar University, Merkez Kamp\"{u}s, K\"{u}tahya, Turkey,

\emph{$^{4)}$} Physics Department, Faculty of Sciences and Arts, Gazi University, 06500 Teknikokullar, Ankara,Turkey,

\emph{$^{5)}$} Institute of Physics, Academy of Sciences, H. Cavid Avenue 33, Baku, Azerbaijan
\end{center}

\vspace{.2cm}
\begin{center}
\emph{$^{}$}Corresponding authors: $^{*)}$ciftci@science.ankara.edu.tr, $^{\dagger)}$hyildiz@dumlupinar.edu.tr
\end{center}

\vspace{.3mm}
\begin{abstract}
Possible single productions of the fourth standard model family charged leptons via anomalous interactions at the future ep 
colliders are studied. Signatures of such anomalous processes and backgrounds are discussed in detail.
\end{abstract} 

\vspace{.3cm}

\vspace{.9cm}
\textbf{1. Introduction}

\vspace{.6cm}
A new era in High-Energy Physics will be opened with the construction of TeV scale colliders {[}1{]} in coming years. The first 
of the such colliders is the Large Hadron Collider (LHC) with 14 TeV center of mass energy and $10^{34}$ cm$^{-2}$s$^{-1}$ design 
luminosity. As for the lepton colliders, there are two realistic proposals namely International Linear Collider (ILC) with 0.5 TeV 
center of mass energy and 2x$10^{34}$ cm$^{-2}$s$^{-1}$ luminosity {[}2{]} and Compact Linear Collider (CLIC) with 1 TeV (second 
stage) center of mass energy and 2.7x$10^{34}$ cm$^{-2}$s$^{-1}$ luminosity {[}3{]}. Construction of ILC or CLIC tangential to the 
LHC ring will give opportunity to build multi-TeV center of mass energy lepton-hadron colliders.

The latest lepton-hadron collider was HERA with $\sqrt{s}\approx 0.3$ TeV. One could built the next lepton-hadron collider by
adding an electron ring at LHC tunnel {[}4{]}. In our opinion even though it is feasible, to built linac-ring type ep collider
is better because of its versatility (possibility of building $\gamma$p, $\gamma$-nucleus and FEL$\gamma$-nucleus colliders). 
Depending on its center of mass energy lepton-hadron colliders (ILC or CLIC on LHC) are named as QCD-Explorer with $\sqrt{s}= 1.4$
TeV {[}5, 6{]} and Energy Frontier-ep collider with $\sqrt{s}$ = 3.74 TeV {[}7{]} (1 fb$^{-1}$ integrated luminosity for both
options). QCD-Explorer is a necessary machine for adequate interpretation of LHC data. Differing from ring-ring type machine,
one could increase the center of mass energy of the collider by increasing the length of the linac (adding more units of ILC
or CLIC accelerating structures). Eventually, it is possible to reach frontier energies by using full size of the ILC or CLIC.
TeV scale hadron, lepton and lepton-hadron colliders will be complementary each other from the view point of physics at TeV 
scale {[}1, 5{]}. For this reason different new phenomena should be studied comparatively in these colliders. LHC, ILC and CLIC 
are well developed with respect to physics potential. However, there is little study on search potential of ep colliders. 
Therefore, they deserve further attention.

It is known that flavor democracy hypothesis forces the existence of the fourth standard model (SM) family {[}8-10{]}. The 
masses of the fourth family fermions are expected to be nearly degenerate and lie between 300 and 700 GeV {[}11{]}. Due to 
large mass value, one could expect a serious contribution of anomalous interactions for the fourth SM family. This paper is 
devoted to detailed study the anomalous single production of the fourth family charged leptons at above ep colliders.\\[.5cm]

\textbf{2. Anomalous Interactions of the Fourth SM Family Lepton}

\vspace{.6cm}
The effective lagrangian for the anomalous interactions of the fourth family charged lepton can be adopted from {[}12{]} with minor
modifications as:

\begin{equation}
L=\left(\frac{K^{l_{i}}_\gamma}{\Lambda}\right)e_{l}g_{e}\overline{l_{4}}\sigma_{\mu\nu}l_{i}F^{\mu\nu}+\left(\frac{K^{l_{i}}_Z}{2\Lambda}\right)g_{Z}\overline{l_{4}}\sigma_{\mu\nu}l_{i}Z^{\mu\nu}+h.c. \hspace{2mm}  ,  \hspace{3mm} (\textit{i}=1,2,3) \\
\end{equation}\\[.1cm]
where $K^{l}_{\gamma,Z}$ are the anomalous couplings for the neutral currents with a photon and a Z boson, respectively. $\Lambda$ 
is the cutoff scale for the new physics and $e_l$ is the lepton charge; $g_{e}$ and $g_{Z}$ are the electroweak coupling 
constants, $g_{Z}=g_{e}/cos\theta_{W}sin\theta_{W}$, where $\theta_{W}$ is the Weinberg angle. In the Eq. (1) 
$\sigma_{\mu\nu} = i(\gamma_{\mu}\gamma_{\nu}-\gamma_{\nu}\gamma_{\mu})/2$. $F^{\mu\nu}$ and $Z^{\mu\nu}$ are field strength 
tensors of the photon and Z boson, respectively. Obviously new interactions will lead to additional decay channels of the fourth 
family leptons. In order to calculate decay widths, we have implemented the new interaction vertices into the CompHEP {[}13{]}. 
Results of the calculations for $l_4$ assuming  $(K/\Lambda)$ = 1 TeV$^{-1}$ are given in Table 1. Let us remind that decay widths 
are proportional to $(K/\Lambda)^2$.

At the same table, we present the widths of the $l_4$ SM decay modes, as well. While calculating these values, we have used
leptonic CKM Matrix elements from {[}14{]}, namely: $V_{\nu_4\tau}$ = 2.34$\times10^{-5}$, $V_{{\nu_4}\mu}$ = 6.81$\times10^{-4}$,
$V_{{\nu_4}e}$ = 4.64$\times10^{-4}$; $V_{{l_4}\nu_\tau}$ = 8.14$\times10^{-4}$, $V_{{l_4}\nu_\mu}$ = 1.28$\times10^{-4}$,
$V_{{l_4}\nu_e}$ = 6.43$\times10^{-6}$. Let us remind that SM decay widths are proportional to $|V_{l_4l\prime}|^2$. Depending on
relative magnitudes of $(K/\Lambda)$ and $|V_{l_4l\prime}|$, SM or anomalous decays will dominate. The total decay width $\Gamma$ 
of the fourth family leptons and the relative branching ratios are given in Fig. 1.

\vspace{ .2cm}

\begin{figure}
\begin{center}
\includegraphics{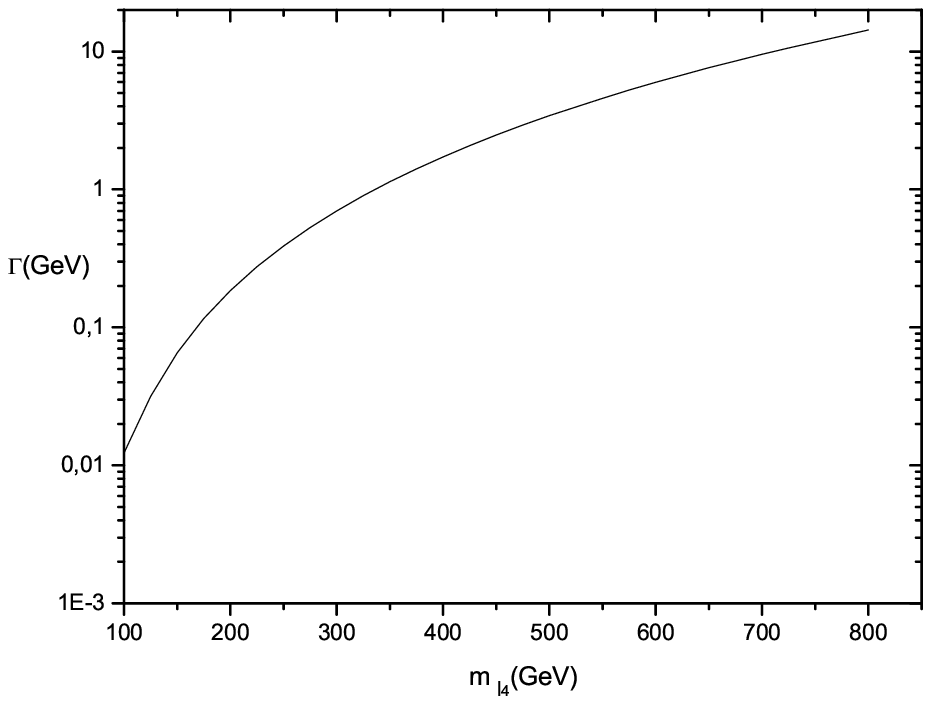}
\includegraphics{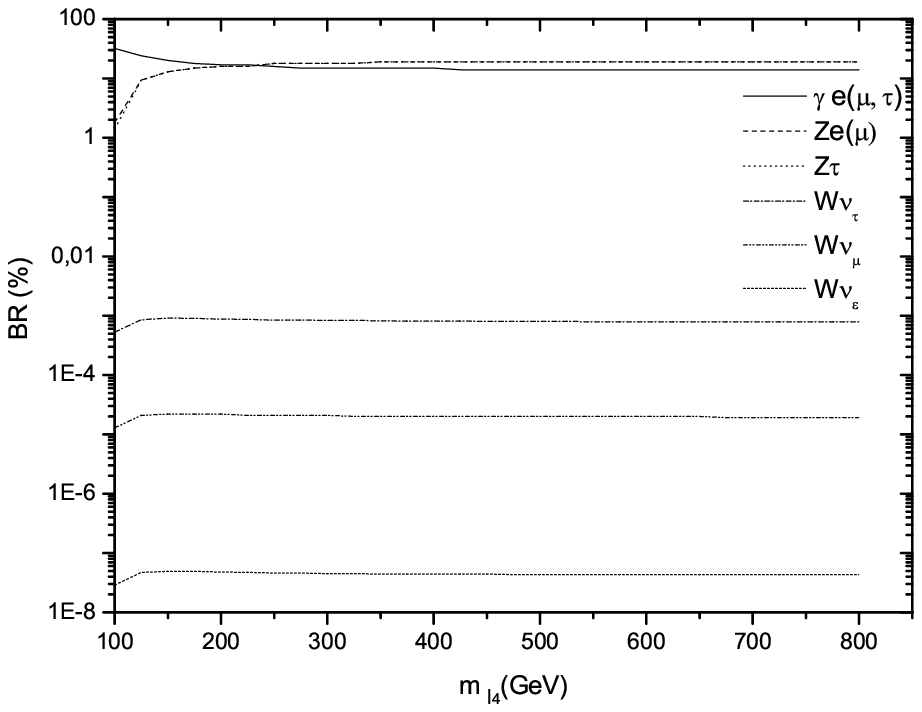}
\end{center}
\caption{(a) The total decay width $\Gamma$ in GeV of the fourth family lepton and (b) the branching ratios (\%) depending on the 
mass of the fourth family lepton.}
\label{fig:Total_width}
\end{figure}

\vspace{ .2cm}
\begin{small}
\begin{center}
\begin{tabular}{l}
Table 1. Branching ratios $(\%)$ and total decay widths for $m_{l_4}$(GeV) \\
\hline
\hline
$m_{l_4}$ \hspace{1.2mm} $\gamma e ({\mu},{\tau})$\hspace{1.8mm} $Z\tau(e,\mu)$ \hspace{1.5mm} W$\nu_\tau(\times10^{-4})$ \hspace{1.7mm}W$\nu_\mu(\times10^{-5})$ \hspace{1.3mm} W$\nu_e(\times10^{-8})$ \hspace{1.2mm} $\Gamma_{Tot}$(GeV) \\
\hline
\hspace{.01mm}100 \hspace{8mm}32 \hspace{7mm}1.4(1.8) \hspace{7mm} 5.3\hspace{13mm} 1.3 \hspace{15mm} 2.9 \hspace{15mm}0.01 \\
\hspace{.01mm}150 \hspace{7mm} 20 \hspace{7mm} 13 \hspace{13mm} 9.1\hspace{13.5mm} 2.2 \hspace{15mm} 4.9 \hspace{14mm} 0.06 \\
200 \hspace{7mm} 17 \hspace{7mm} 16 \hspace{13mm} 8.8 \hspace{12.5mm} 2.2 \hspace{15mm} 4.8 \hspace{14mm} 0.18 \\
300 \hspace{7mm} 15 \hspace{7mm} 18 \hspace{13mm} 8.3\hspace{14mm} 2.1 \hspace{15mm} 4.5 \hspace{14mm} 0.70 \\
700 \hspace{7mm} 14 \hspace{7mm} 19 \hspace{13mm} 7.9\hspace{14mm} 1.9 \hspace{15mm} 4.3 \hspace{14mm} 9.54 \\
\hline
\hline
\end{tabular}
\end{center}
\end{small}

\vspace{ .6cm}

\textbf{3. Anomalous Single Production of the Fourth SM Family Leptons at ep Colliders}

\vspace{ .5cm}
In ep colliders ($\sqrt{s}$ = 1.4 TeV and $\sqrt{s}$ = 3.74 TeV are considered), we investigate single anomalous production of 
the fourth SM family leptons. The calculated cross-sections for $l_4$ are displayed in Fig. 2 and Fig. 3. We consider 
$ep\rightarrow l_4 X \rightarrow \gamma\mu X$ and $ep\rightarrow l_4 X \rightarrow ZeX$ processes. In the former process 
anomalous decay of $l_4$ can be detected easily at ep colliders due to no background. Number of events for $m_{l_4}$ = 300 (700) 
GeV at $\sqrt{s}$ = 1.4 TeV and $\sqrt{s}$ = 3.74 TeV with 1 $fb^{-1}$ integrated luminosity are 240 (30) and 1170 (460),
respectively.

However, the latter process has SM backgrounds. In order to extract the fourth generation lepton signal and to suppress the 
background, we impose cuts on the eZ invariant mass. Thus, in calculations, we used $|m_{eZ}-m_{l_{4}}|<$25 
GeV for the mass range $m_{l_{4}}=100-1000$ GeV and $|m_{eZ}-m_{l_{4}}|<$50 GeV for the mass range of $1-2.6$ TeV with 
$p_{T}^{q}>$10 GeV cuts. All calculations for the signal and background were done by using the high-energy package CompHEP 
with CTEQ6L {[}15{]} distribution function. We present signal and background cross-sections with respect to invariant 
mass of eZ in Tables 2 and 3 for $\sqrt{s}$ = 1.4 TeV, $\sqrt{s}$ = 3.74 TeV options, respectively.

\vspace{ .4cm}
In order to examine the potential of the collider to search for the investigation of the anomalous interactions of the fourth SM 
family lepton in detail, we define statistical significance (SS) of the signal:
\vspace{ .1cm}
\begin{equation}
SS=\dfrac{\left | \sigma_{S+B}-\sigma_{B}\right |}{\sqrt\sigma_B}\sqrt L_{int}
\end{equation}\\[.1cm]where $L_{int}$ is the integrated luminosity of the collider. The values SS evaluated at each fourth SM family 
lepton mass points are shown in the last column of Table 2 and Table 3. As seen from tables, one obtains good SS for the expected mass 
range of leptons. \\
\vspace{ .05mm}

\begin{figure}
\begin{center}
\includegraphics{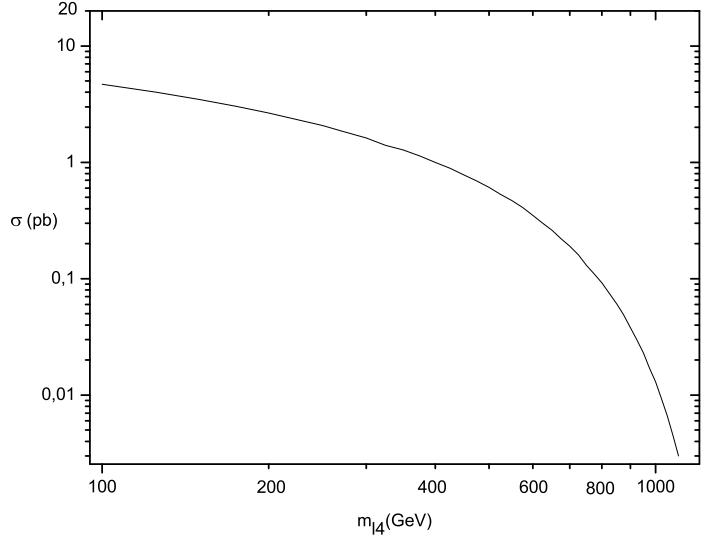}
\end{center}
\caption{The total production cross-section of the process $ep\rightarrow l_4$X as a function of $m_{l_4}$ with the center of
mass energy $\sqrt{s}$ = 1.4 TeV.}
\label{fig:ep14}
\end{figure}

\begin{figure}[hbtp]
\begin{center}
\includegraphics{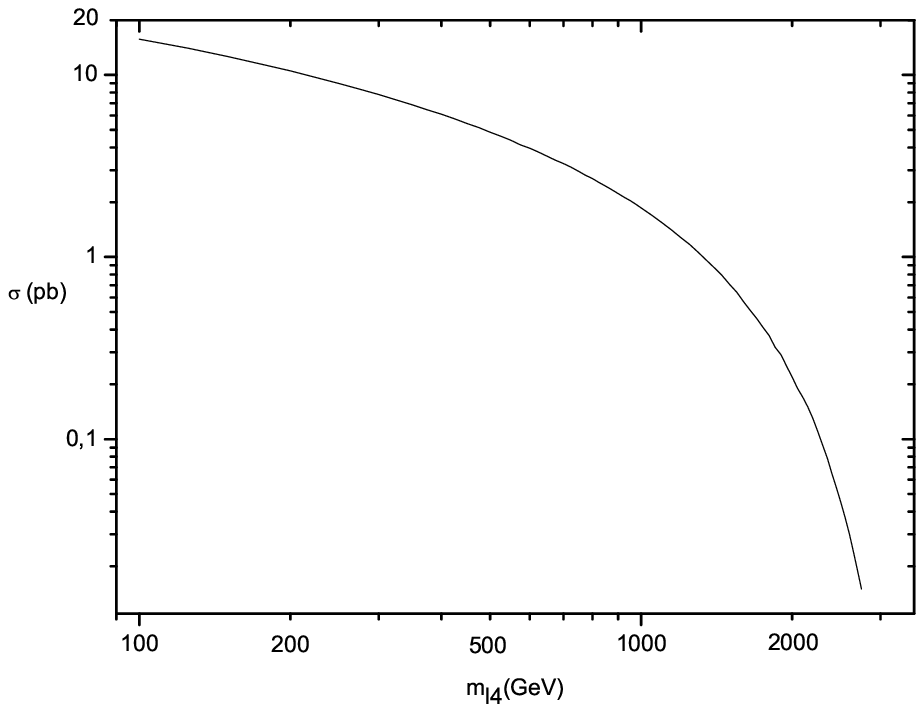}
\end{center}
\caption{The total production cross-section of the process $ep\rightarrow l_4$X as a function of $m_{l_4}$ with the center of
mass energy $\sqrt{s}$ = 3.74 TeV.}
\label{fig:ep374}
\end{figure}

\vspace{ .1mm}
\begin{small}
\begin{center}
\begin{tabular}{l}
Table 2. Total cross section of signal and background \\ \hspace{12.5mm} for $\sqrt{s}$ = 1.4 TeV and $(K/\Lambda)$ = 1TeV$^{-1}$ \\
\hline
\hline
$m_{l_4}$(GeV) \hspace{10mm} $\sigma_{S+B}$(pb) \hspace{10mm} $\sigma_B$(pb) \hspace{9mm} SS  \\
\hline
100 \hspace{22mm}0.13 \hspace{18mm}0.1 \hspace{15.5mm} 3 \\
200 \hspace{21mm} 0.49 \hspace{17mm} 0.09 \hspace{12mm} 42.16 \\
300\hspace{22mm} 0.33 \hspace{17mm} 0.05 \hspace{12mm} 39.60 \\
400\hspace{22mm} 0.21 \hspace{17mm} 0.03 \hspace{12mm} 32.86 \\
500\hspace{22mm} 0.13 \hspace{17mm} 0.02 \hspace{12mm} 24.59 \\
600\hspace{22mm} 0.075 \hspace{15mm} 0.015 \hspace{10.5mm} 15.49 \\
700\hspace{22mm} 0.039 \hspace{15mm} 0.009 \hspace{10.5mm} 10.00 \\
800\hspace{22mm} 0.019 \hspace{15mm} 0.005 \hspace{12mm} 6.26 \\
900\hspace{22mm} 0.0076 \hspace{13.5mm} 0.002 \hspace{12mm} 3.96 \\
1000\hspace{20.3mm} 0.0028 \hspace{13.5mm} 0.0011 \hspace{10mm} 1.62 \\
\hline
\hline
\end{tabular}
\end{center}
\end{small}

\vspace{ .5cm}
Fig. 4 shows the invariant mass $m_{eZ}$ distribution in the reaction $ep\rightarrow l_4 X \rightarrow ZeX$ for the SM background 
and the signal with the inclusion of the fourth family lepton with masses $m_{l_4}$ = 300 GeV, $m_{l_4}$ = 500 GeV, and 
$m_{l_4}$ = 700 GeV for QCD Explorer.

Fig. 5 shows the invariant mass $m_{eZ}$ distribution in the reaction $ep\rightarrow l_4 X \rightarrow ZeX$ for the SM background 
and the signal  with the inclusion of the fourth family lepton with masses $m_{l_4}$ = 500 GeV, $m_{l_4}$ = 1 TeV, and 
$m_{l_4}$ = 1.5 TeV for Energy Frontier-ep collider.

\vspace{ .3cm}

\begin{small}
\begin{center}
\begin{tabular}{l}
Table 3. Total cross section of signal and background for \\ \hspace{12.5mm} for $\sqrt{s}$ = 3.74 TeV, $(K/\Lambda)$ = 1TeV$^{-1}$ \\
\hline
\hline
$m_{l_4}$(GeV) \hspace{10mm} $\sigma_{S+B}$(pb) \hspace{10mm} $\sigma_B$(pb) \hspace{11mm} SS  \\
\hline
200 \hspace{22mm}1.234 \hspace{16mm}0.064 \hspace{12mm} 146.25 \\
400 \hspace{21mm} 0.856 \hspace{15mm} 0.026 \hspace{12mm} 162.77 \\
600\hspace{22mm} 0.554 \hspace{15mm} 0.014 \hspace{12mm} 144.32 \\
800\hspace{22mm} 0.339 \hspace{15mm} 0.009 \hspace{12mm} 110.00 \\
1000\hspace{20.5mm} 0.32 \hspace{16.5mm} 0.03 \hspace{15.1mm} 52.94 \\
1200\hspace{20.5mm} 0.19 \hspace{16.5mm} 0.02 \hspace{15.1mm} 38.01 \\
1400\hspace{20.5mm} 0.105 \hspace{15mm} 0.015 \hspace{13.5mm} 23.24 \\
1600\hspace{20.5mm} 0.062 \hspace{15mm} 0.012 \hspace{13.5mm} 14.43 \\
1800\hspace{20.5mm} 0.028 \hspace{15mm} 0.008 \hspace{15mm} 7.07 \\
2000\hspace{20.5mm} 0.016 \hspace{15mm} 0.006 \hspace{15mm} 4.08 \\
2200\hspace{20.5mm} 0.009 \hspace{15mm} 0.004 \hspace{15mm} 2.5 \\
2400\hspace{20.5mm} 0.004 \hspace{15mm} 0.002 \hspace{15mm} 1.41 \\
2600\hspace{20.5mm} 0.002 \hspace{15mm} 0.001 \hspace{15mm} 1 \\
\hline
\hline
\end{tabular}
\end{center}
\end{small}

\vspace{ .2cm}

\begin{figure}
\begin{center}
\includegraphics{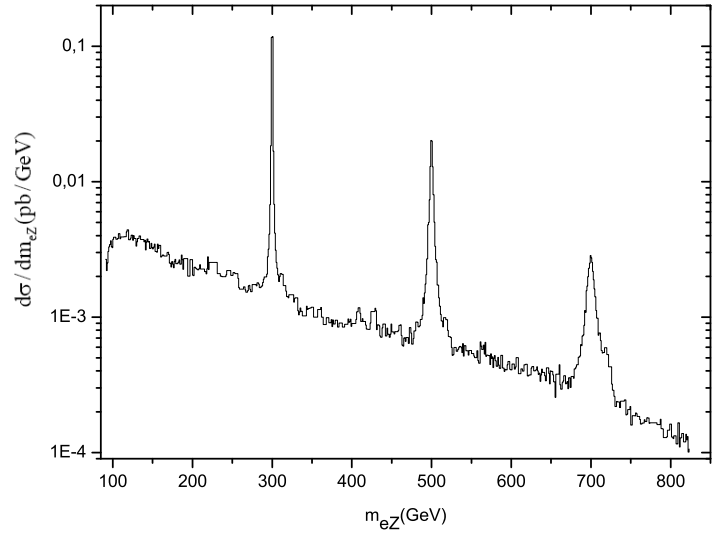}
\end{center}
\caption{Invariant mass $m_{eZ}$ distribution of signal and background at QCD Explorer.}
\label{fig:Sig+back_1.4}
\end{figure}

\begin{figure}
\begin{center}
\includegraphics{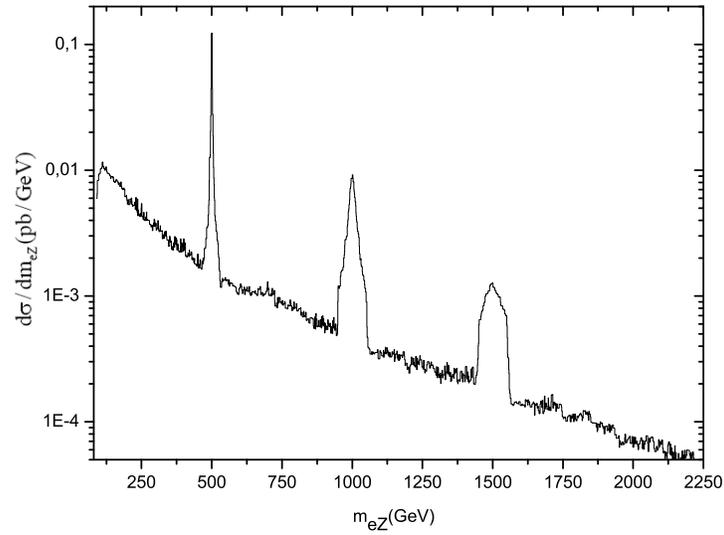}
\end{center}
\caption{Invariant mass $m_{eZ}$ distribution of signal and background at Energy Frontier-ep collider.}
\label{fig:Sig+back_3.74}
\end{figure}

\vspace{ .2cm}

\textbf{4. Conclusion}

\vspace{.3cm}
As a result of study it is seen that ep colliders are advantageous to investigate anomalous interactions of the fourth SM family
charged lepton. QCD Explorer will cover fourth family charged lepton up to 900 GeV while at Energy Frontier-ep collider the 
masses up to 2100 GeV can be reached. The regions are similar to reaches for anomaluos production of the fourth family neutrino 
{[}16{]}. Hadron colliders including LHC are not very good place to investigate fourth family leptons {[}17{]} whereas, the 
capacity of lepton colliders are limited by their center of mass energy.

\vspace{ .4cm}

\textbf{Acknowledgments}

\vspace{.3cm}
This work is supported by TAEK and DPT with grant number DPT2006K-120470. One of the authors (H. Duran Y{\i}ld{\i}z) supported by T\"{U}B\^ITAK with 
the project number 105T442.

\vspace{ .2cm}

\end{document}